\documentstyle[prl,manuscript,aps]{revtex}
\def\figsize{12cm}

\begin{document}
\draft

\input epsf.sty

\title{Scale Invariance in the Nonstationarity 
of Physiological Signals}

\author{Pedro Bernaola-Galv\'{a}n$^{1,2}$, Plamen~Ch.~Ivanov$^{1,3}$,
Lu\'{\i}s~A.~Nunes~Amaral$^{1,3}$, \\
Ary~L.~Goldberger$^3$, H.~Eugene~Stanley$^1$}

\address{ $^1$ Center for Polymer Studies and Department of Physics,
  		Boston University, Boston, MA 02215\\
          $^2$ Departamento de F\'{\i}sica Aplicada II, Universidad de
                M\'alaga E-29071, Spain\\
	  $^3$ Cardiovascular Division, Harvard Medical School,
		Beth Israel Deaconess Medical Center, Boston, MA 02215}

\date{\today}

\maketitle
\begin{abstract}

We introduce a segmentation algorithm to probe temporal organization
of heterogeneities in human heartbeat interval time series.  We find
that the lengths of segments with different local values of heart
rates follow a power-law distribution. This scale-invariant structure
is not a simple consequence of the long-range correlations present in
the data. We also find that the differences in mean heart rates
between consecutive segments display a common functional form, but
with different parameters for healthy individuals and for patients
with heart failure. This finding may provide information into the way
heart rate variability is reduced in cardiac disease.

\end{abstract}
\bigskip

\newpage

\begin{figure}
\centerline{
  \epsfxsize=\figsize {\epsfbox{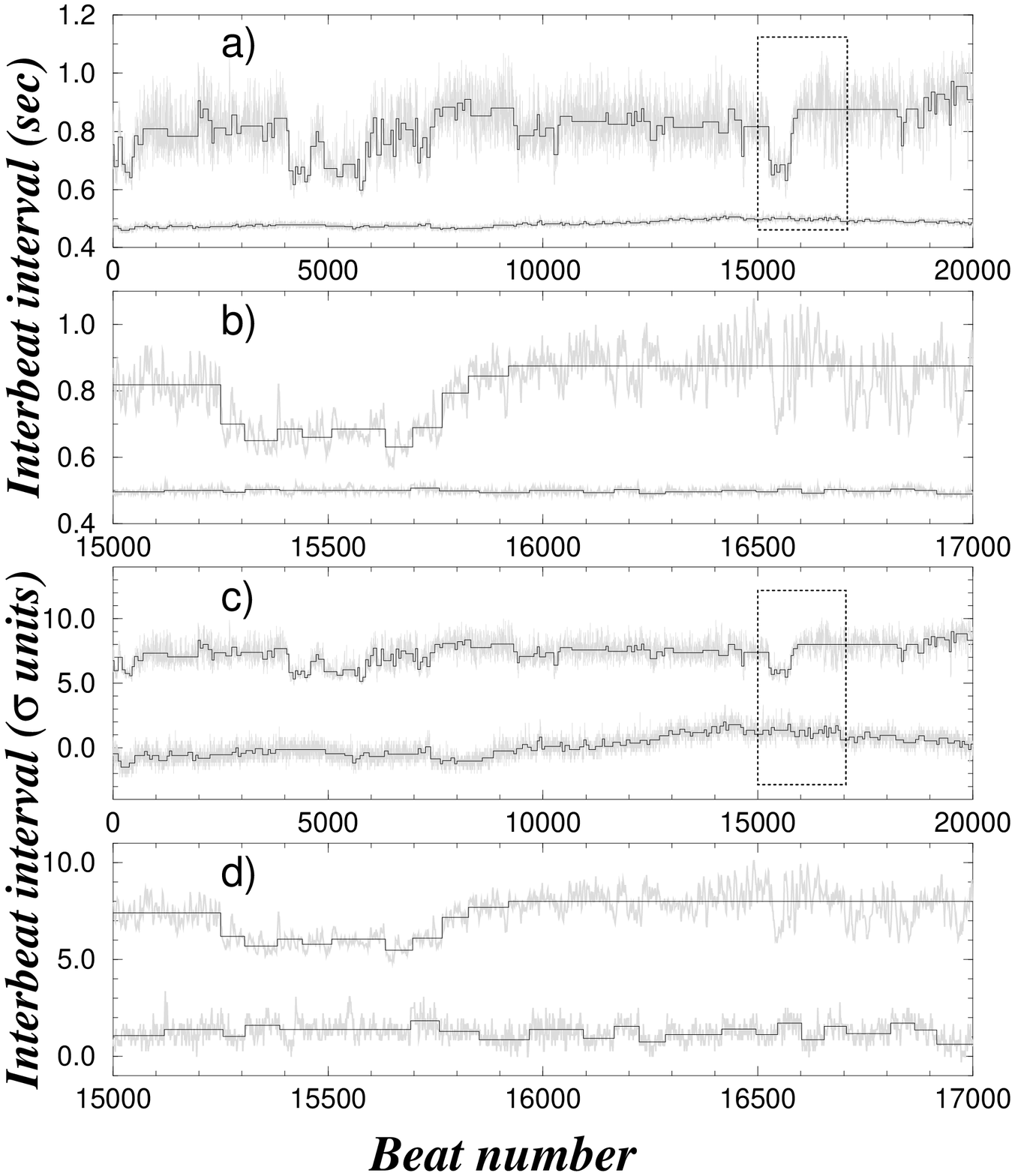}}
  \vspace{0.5cm}
}
\caption{(a) Plot of 20,000 interbeat intervals ($\approx$ 6 hours) for 
a healthy subject (upper curve) and a subject with heart failure (bottom curve).
Note the larger variability and patchiness for the healthy record.
(b) Magnification of a small fraction (2000 beats) of the signals in (a). 
(c) Same signals as displayed in (a) after subtracting the
global average  and dividing by the global standard deviation;
after this normalization both signals appear very similar.
(d) Magnification of a small fraction (2000 beats) of the signals in (c).}
\label{fig1}
\end{figure}

A time series is 
stationary if the mean, standard deviation and all higher moments, as well
as the correlation functions, are invariant under
time translation \cite{DefStat}.
Signals that do not obey these conditions are nonstationary.
Nonstationarity is a prominent feature of biological variability that can be
associated with regimes (segments) of different statistical properties.
The borders between different segments can be gradual or abrupt
(Fig. \ref{fig1}).

A major problem in contemporary physiology
is the presence of nonstationarity in time series
generated under free-running conditions \cite{Variability}.
Physiological signals
obtained under widely-varying conditions raise serious
challenges to both technical and fundamental aspects of time series
analysis.
By filtering out effects of nonstationarity, much work has focused on
``intrinsic properties'' of physiological signals \cite{FilterOut}.
This approach is based on the
implicit assumption that the nonstationarity 
arises simply from changes in environmental conditions ---
e.g., different daily activities ---
so environmental ``noise'' could be treated as a
``trend'' and distinguished from the more subtle fluctuations that may
reveal intrinsic correlation properties of the dynamics.
Indeed, important scale-invariant features in physiological processes
were recently revealed after filtering out
masking effects of nonstationarity \cite{FractalHeart}.
However, nonstationarity itself is also an important feature of physiological
time series and is known to change from healthy to pathological conditions
\cite{Nonstat}, suggesting more than only enviromental conditions are 
reflected in the phenomena.
Thus one would expect that there is a non-trivial structure associated with
the nonstationarity in physiological signals, which may change with
disease.
To test this hypothesis we focus on one statistical property,
the mean heart rate, which is related to physiologic responses and
is commonly used for medical evaluation.

The problem is to partition a nonstationary time series, which is composed
of many segments with different mean value, in such a way as to
maximize the difference in the mean values between adjacent segments.
We apply the following procedure: we move a
sliding pointer from left to right along the signal.
At each position of the pointer, we compute the mean of the subset of the
signal to the left of the pointer ($\mu_{\rm left}$) and to the right
($\mu_{\rm right}$). To measure the difference between $\mu_{\rm left}$ and
$\mu_{\rm right}$, we compute the t-statistic \cite{NumRec}:
\begin{equation}
t \equiv \left | \frac{\mu_{\rm left}-\mu_{\rm right}}{s_D} \right |
\label{student}
\end{equation}
where $s_D$ is the pooled variance \cite{pooled}.

\begin{figure}
\centerline{
  \epsfxsize=\figsize {\epsfbox{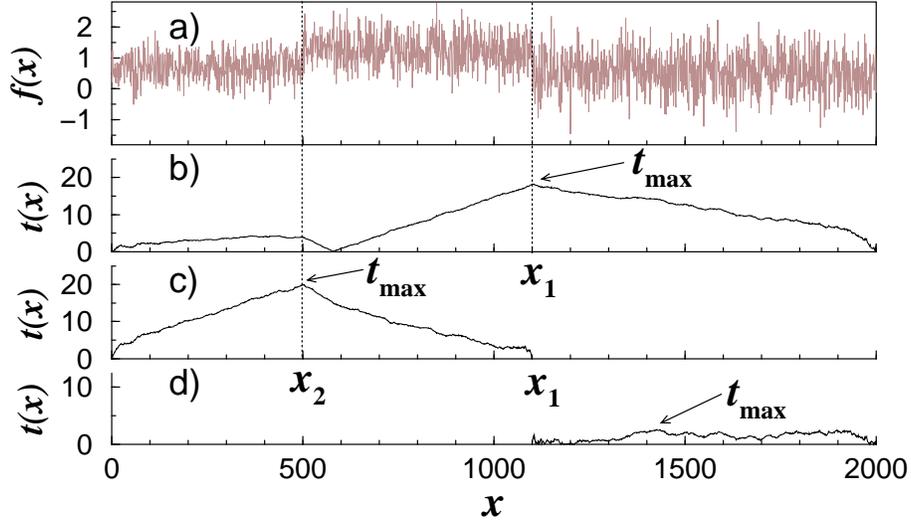}}
}
\caption{(a) An artificial time series $f(x)$ composed of three segments
with different mean values. (b) Values of the statistic $t$, defined in
Eq.~(\protect\ref{student}), obtained by moving the pointer along the
time series. Note that $t_{\rm max}$ is reached at $x_1$. We find that if
${\cal P}(t_{\rm max}) \geq {\cal P}_0=95\%$, and so we cut the series
at $x_1$. (c) We iterate the procedure with the segment $[0,x_1]$.
We find that
${\cal P}(t_{\rm max}) \geq 95\%$ and we also find  that the significance
of $t$ computed between $[x_2,x_1]$ and $[x_2,2000]$ is greater than 
$95\%$, so the series is cut at $x_2$. (d) We iterate the procedure with
the segment $[x_1,2000]$. Now, ${\cal P}(t_{\rm max}) \leq 95\%$, so this
segment is not cut.}
\label{figExample}
\end{figure}

We next determine the position of the pointer for which $t$ reaches its
maximum value, $t_{\rm max}$, and compute the statistical significance of
$t_{\rm max}$ \cite{siglevel}. We check if this significance exceeds a
given threshold ${\cal P}_0$. If so, then the signal is cut at
this point into two subsequences; otherwise
the signal remains undivided. If the sequence is cut, the procedure
continues recursively for
each of the two resulting subsequences created by each cut.
Before a new cut is accepted,
we also compute $t$ between the right-hand
new segment and its right neighbor (obtained by
a previous cut) and the $t$ between the left-hand new segment and
its left neighbor (also obtained by a previous cut) and check if both
values of $t$ have a statistical significance exceeding ${\cal P}_0$.
If so, we proceed with the new cut; otherwise we do not cut.
This ensures that all resulting segments have a statistically significant
difference in their means.
The process stops when
none of the possible cutting points has a significance exceeding
${\cal P}_0$,
and we say that the signal has been segmented at the ``significance
level ${\cal P}_0$'' (Fig.~\ref{figExample}).

Our method leads to partitioning of a time series into segments with
well-defined means, each significantly different from the mean of the adjacent
segments (Fig. \ref{fig1}). This allows us to probe the
nonstationarity in a signal through the statistical analysis of the
properties of the segments.

Here we consider 47 datasets
from 18 healthy subjects, 17 records of cosmonauts during orbital flight
and 12 patients with congestive heart failure \cite{MIT}.
We separately analyze 6--hour
long subsets of each dataset, corresponding to the periods when the
subject is awake or sleeping. Figure \ref{fig1} shows a representative dataset
of a healthy subject, and a subject with heart failure. Superposed on the
interbeat interval series, we also plot the segments obtained by means of our
segmentation algorithm.

To quantify the nonstationarity in heart rate variability, we study the
statistical properties of the segments corresponding to parts of the signal
with significantly different mean values. To characterize the segments,
we analyze two quantities:
(i) the length of the segments;
(ii) the absolute values of the differences between the mean values
of consecutive segments, which we call jumps.

\noindent (i) {\sl Distribution of segment lengths ---}
Healthy subjects typically exhibit nonstationary behavior associated
with large variability, trends, and segments
with large differences in their mean values, while data from heart failure
subjects are characterized by reduced variability and appear to be more homogeneous
(Fig. \ref{fig1})
 \cite{Nonstat}.
Thus, one might naively expect that signals from healthy subjects will be
characterized by a large number of segments,
while signals from heart failure subjects will exhibit a smaller number of segments
(i.e., the average length of the segments for healthy subjects could be
expected to be smaller than for heart failure subjects).

\begin{figure}
\centerline{
  \epsfxsize=\figsize {\epsfbox{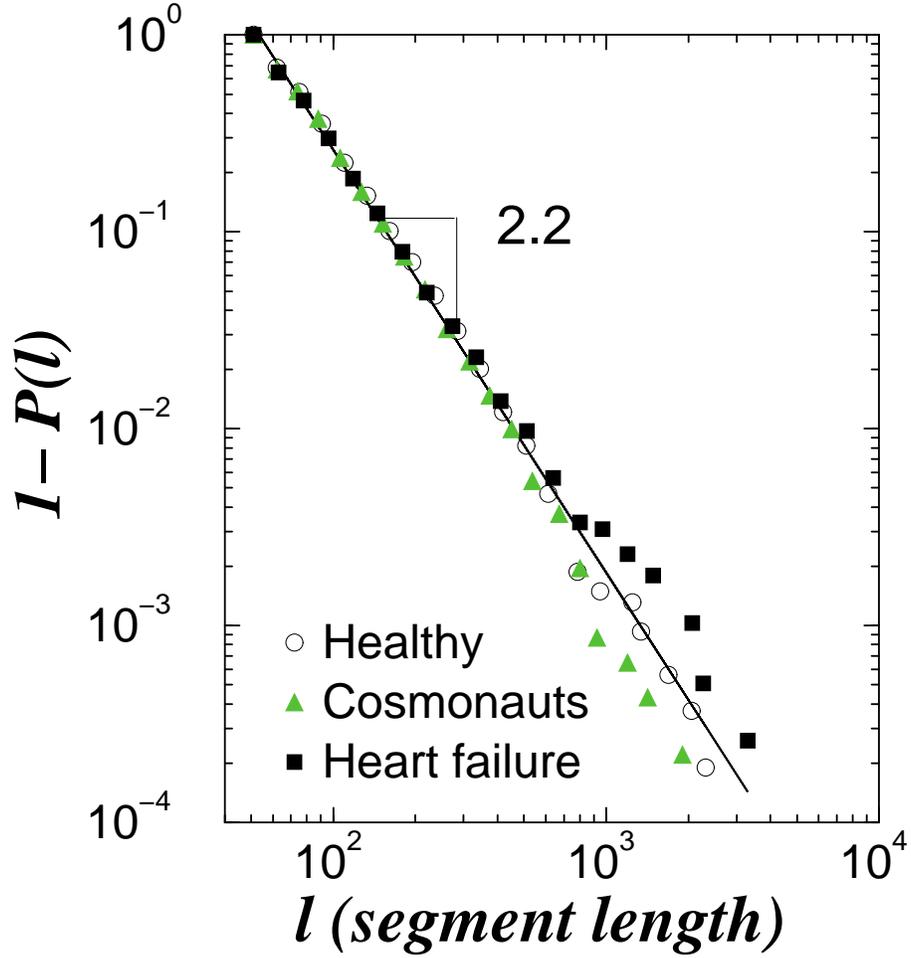}}
}
\caption{
Probability of finding segments with a length $\ell$ larger than a given
value for the segments obtained from all subjects in the healthy, cosmonauts
and heart failure groups during during daily activity.
The significance level is fixed to $\protect {\cal P}_0=95 \% $, and the
imposed minimum length of the segments is $\protect\ell_0=50$ beats. 
For all three groups we find a power law in the distribution of segment
lengths with exponent $\beta \approx 2.2$,
and we find that $\beta$
depends on $\ell_0$ and on ${\cal P}_0$.
For all $\ell_0$ and ${\cal P}_0$, the value of $\beta$ is the same for each
three groups.}
\label{fig2}
\end{figure}

To test this hypothesis, we apply the segmentation algorithm to
6--hour records of interbeat intervals during daily activity, and find
that for each healthy subject the distribution of segment lengths is
well described by a power law with an identical exponent, indicating
absence of a characteristic length for the segments.
Surprisingly, we find that this power law remains unchanged for
records obtained from cosmonauts during orbital flight (under
conditions of microgravity) and for patients with heart failure
(Fig. \ref{fig2}).  A similar common type of
behavior is also observed from 6--hour records during
sleep for all three groups \cite{night}.

\begin{figure}
\centerline{
  \epsfxsize=\figsize {\epsfbox{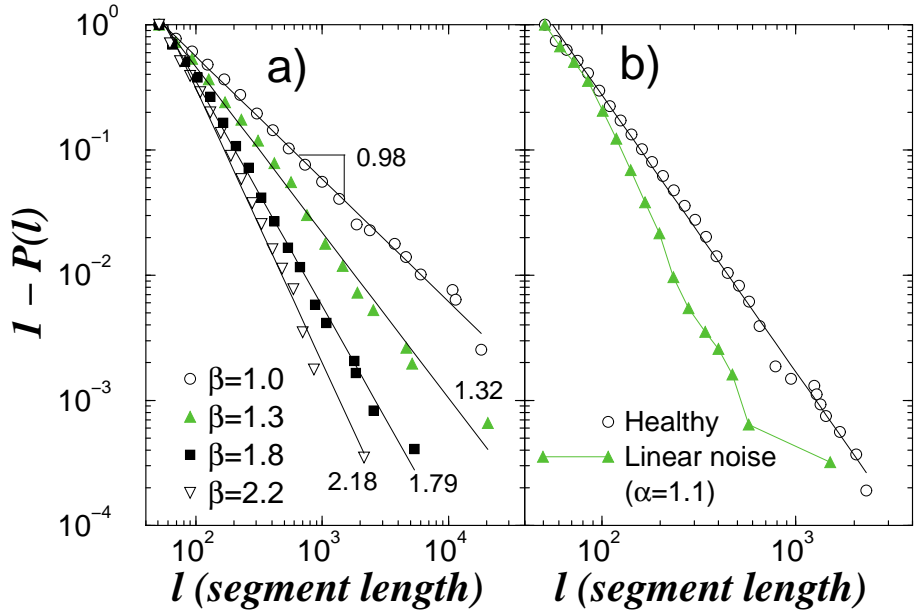}}
}
\caption{
(a) Testing the validity of the observed 
power-law behavior in the distribution of segment lengths.
We generate a surrogate signal formed by joining segments
of white noise with standard deviation $\sigma = 0.5$ and average values chosen
randomly from the interval $[0,1]$. We chose the lengths of these segments
from a power-law distribution with a given exponent $\beta$.
The test shows that the distributions
obtained after segmenting the surrogate data generated from  power-law
distributions
with nearby values of their exponents appear clearly separated. This suggests
that the distributions for the healthy,
cosmonauts and congestive heart failure subjects in Fig.~{\protect\ref{fig2}}
are indeed identical.
(b) Testing the effect of correlations in the heartbeat fluctuations
on the segmentation. We generate 10 realizations, each with length of 26,000 points,
of a linear Gaussian-distributed correlated noise with an exponent $\alpha=1.1$
{\protect\cite{GenNoise}}.
This exponent is calculated using the detrended fluctuation analysis method
and is identical to the exponent $\alpha$ observed for the heartbeat data
{\protect\cite{Peng93}}.
The distribution of segment lengths for this correlated noise does not follow
the power law found for the heartbeat data. This test suggests that the observed
scale-invariant behavior in the distributions of segment lengths in the heartbeat
is not simply due to the correlations. According to the results in (a), the
differences found between heartbeat data and correlated noise are significant.
To verify that the curvature found in the distribution of segments
for the noise is not due to finite size effects, we also repeated the test
with longer realizations of the noise (1,000,000).}
\label{fig2bis}
\end{figure}

To verify the results of the segmentation procedure, we perform several tests.
First, we check the validity of the observed power law in the distribution
of segment lengths. We generate a surrogate signal formed by joining segments
of white noise with standard deviation $\sigma = 0.5$, and mean values chosen
randomly from the interval $[0,1]$. We choose the lengths of these segments
from a power-law distribution with a given exponent. Even when the
difference between the mean values of adjacent segments is smaller than
the standard deviation of the noise inside the segments, we find that our
procedure partitions the surrogate signal into segments with lengths that reproduce
the original power-law distribution [Fig. \ref{fig2bis}(a)].
This test shows that the distributions obtained after segmenting surrogate data
with similar values of their exponents, appear clearly different from each other,
making
more plausible that the distributions obtained for the lengths of the segments for
the healthy, cosmonauts and congestive heart failure subjects (Fig.\ref{fig2})
follow indeed an identical distribution.

Second, we test if the observed power-law distribution for the segment lengths
is simply due to the known presence of long-range correlations in the heartbeat
interval series \cite{Peng93}. For that, we generate correlated linear noise
\cite{GenNoise} with the same correlation exponent as the heartbeat data.
We find that the distribution
of segment lengths obtained for the linear noise
differs from the distribution obtained for the heartbeat data
[Fig. \ref{fig2bis}(b)]. For the noise, the distribution
decays faster, which means that these signals are more segmented
than the heart data. In fact, for different linear
noises with a broad range of correlation exponents, we do not find power-law
behavior in the distribution of the segments. Thus we conclude that the linear
correlations are not sufficient to explain the power-law distribution of
segment lengths in the heartbeat data.

\noindent (ii) {\sl Differences between the mean values  of consecutive
segments (jumps) ---} Different healthy records can be characterized by different
overall
variance, depending on the activity and the individual characteristics of
the subjects. Moreover, subjects with heart failure exhibit
interbeat intervals with lower mean and reduced beat-to-beat variability (lower
standard deviation).
Thus one can trivially assume that these larger jumps in healthy records are
due only to the fact that their average standard deviation is larger
[Fig.~\ref{fig1}(a)(b)].
In order to systematically compare the statistical properties
of the jumps between different individuals and different groups, we normalize
each time series by subtracting the global average (over
6 hours) and dividing by the global standard deviation. In this way, all
individual time series have zero mean and unit standard deviation
[Fig.~\ref{fig1}(c)(d)].
Such a normalization does not affect the results of our segmentation procedure.

\begin{figure}
\centerline{
  \epsfxsize=\figsize {\epsfbox{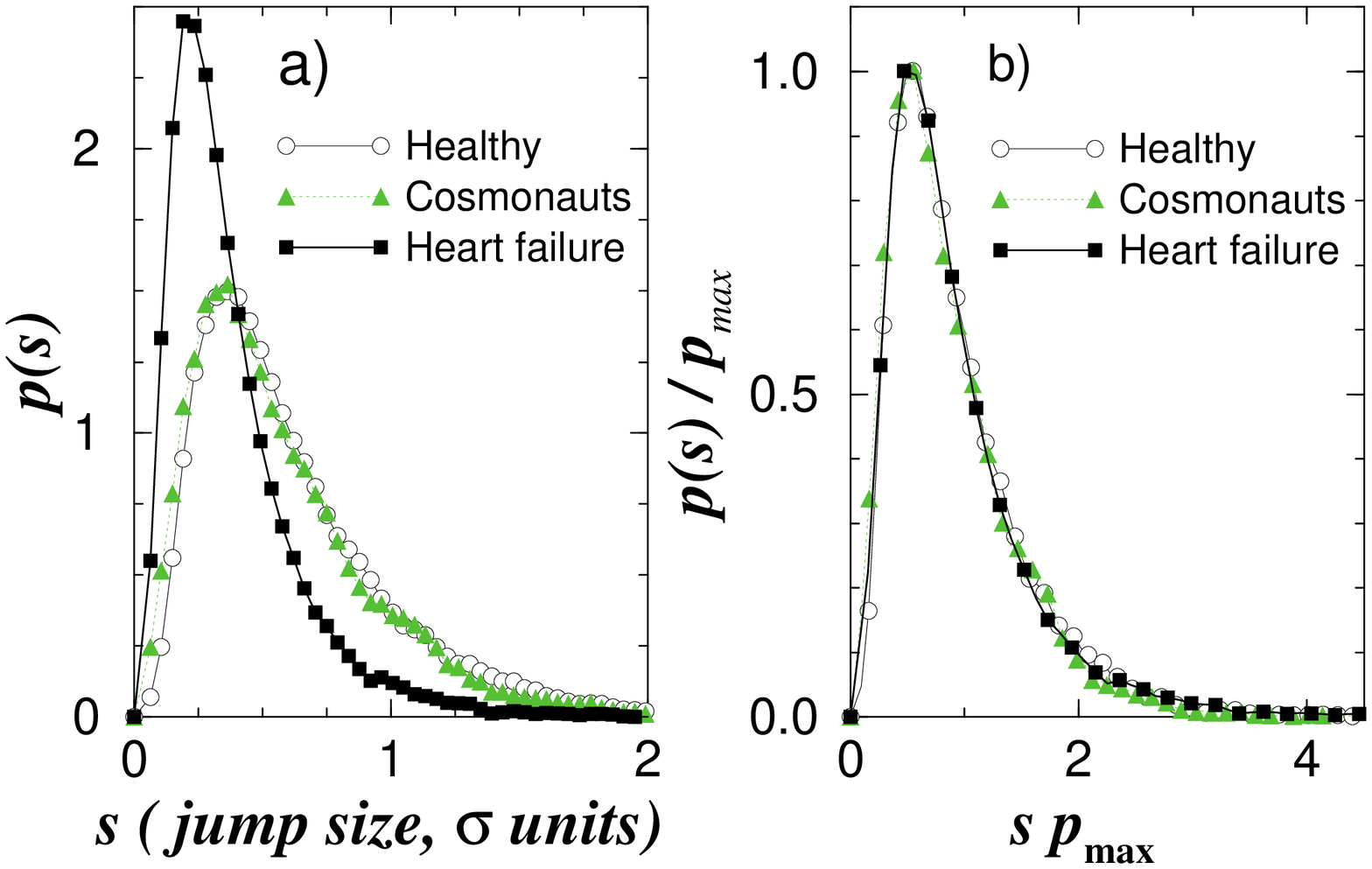}}
}  
\caption{(a) Probability distribution
of the absolute value of the difference between the mean values (`jumps') of
consecutive segments. Both healthy and cosmonaut subjects follow an identical
distribution while the heart-failure subjects follow a quite different
distribution with higher probability for small jumps consistent with reports
of smaller variability in heart failure subjects {\protect\cite{Nonstat}}.
All distributions are normalized to unit area.
(b) Same probability distributions as in (a), after rescaling $P(s)$ by
$P_{\rm max}$, and $s$ by $1/P_{\rm max}$. This homogeneous transformation
preserves the normalization to unit area. The data points collapse onto
a single curve.}
\label{fig3}
\end{figure}

We find that both the healthy subjects and the cosmonauts follow identical
distributions, but the distribution of the jumps obtained from the
heart failure group are markedly different --- centered around lower values ---
indicating that, even after normalization, there is a higher probability for
smaller jumps compared to the healthy subjects [Fig.~\ref{fig3}(a)]. Note
that the distributions for all groups appear to follow an identical homogeneous
functional form,
so we can collapse these distributions on top of each
other by means of a homogeneous transformation [Fig.~\ref{fig3}(b)].
The ratio between the scaling parameters used in this transformation
gives us a factor by which this feature of the heartrate variability is reduced for
the subjects with heart failure as compared to the healthy subjects.
This finding indicates that, although the heartrate variability is reduced with
disease, there may be a common structure to this variability, reflected
in the identical functional form.
These observations agree with previously reported results for the distribution of
heartbeat fluctuations obtained by means of wavelet and Hilbert transforms
\cite{Ivanov96}.

In summary, we present a new method
to probe the nonstationarity of a signal by partitioning it
into segments with different mean values. We find 
a scale-invariant structure in the nonstationarity of a time series
representative of a complex dynamics, namely the human heartbeat. This
structure is characterized by a power-law
distribution of the lengths of segments with a scaling 
exponent which does not change under certain pathological conditions and cannot
be explained by the presence of correlations in the data.
We find also a common structure to the jumps between consecutive segments,
with a change in the scaling parameters with disease.

We thank Y. Ashkenazy, V. Schulte-Frohlinde, I. Grosse, S. Havlin,
S. Mossa, C.-K. Peng, and Z. Struzik for helpful discussions and
suggestions, grants BIO99-0651-CO2-01 (from the Spanish Government) and
NIH/NCRR (P41RR13622), NASA, and the Mathers Charitable Foundation for
support.


\newpage

%
%

\newpage

%

\newpage

%

%

%


\begin{references}


\bibitem{DefStat}
R.L.~Stratonovich, {\sl Topics in the Theory of Random Noise}, vol.1
(Gordon and Breach, New York, 1981).


\bibitem{Variability}
R.I.~Kitney and O.~Rompelman, {\sl The Study of Heart Rate Variability}
(Oxford Univ. Press, London, 1980); 
J.B.~Bassingthwaighte, L.S.~Liebovitch and B.J.~West,
{\sl Fractal Physiology} (Oxford Univ. Press, New York, 1994);
B.J.~West, {\sl Fractal Physiology and Chaos in Medicine} (World
Scientific, Singapore, 1990).


\bibitem{FilterOut} 
H.~Kantz and T.~Schreiber, {\sl Nonlinear Time Series Analysis.}
(Cambridge Univ. Press, Cambridge, 1997);
T.~Schreiber, Phys. Rev. Lett. {\bf 78}, 843 (1997);
A.~Witt, J.~Kurths and A.~Pikovsky, Phys. Rev. E. {\bf 58}, 1800 (1998);
G.~Mayer-Kress, Integ. Physiol. Behav. Sci. {\bf 29}, 205 (1994);
R.~Hegger, H.~Kantz, and L.~Matassini, Phys. Rev. Lett {\bf 84},
3197 (2000).



\bibitem{FractalHeart}
M.~Kobayashi and T.~Musha, IEEE Trans Biomed Eng. {\bf 29}, 456 (1982);
J.M.~Hausdorff {\sl et al.}, J. Appl. Physiol. {\bf 80}, 1448 (1996);
M.F.~Shlesinger, Ann. NY Acad. Sci. {\bf 504}, 214 (1987);
L.S.~Liebovitch, Biophys. J. {\bf 55}, 373 (1989);
A.~Arneodo {\sl et al.}, Physica D {\bf 96}, 291 (1996).



\bibitem{Nonstat}
M.M.~Wolf {\sl et al.}, Med. J. Aust. {\bf 2},
52 (1978);
C.~Guilleminault {\sl et al.}, Lancet {\bf 1}, 126 (1984);
A.L.~Goldberger {\sl et al.}, Experientia {\bf 44}, 983 (1988).



\bibitem{NumRec} W.H. Press {\sl et al.},
{\sl Numerical Recipes in FORTRAN} (Cambridge University Press, 
Cambridge, 1994).

\bibitem{pooled}
The pooled variance is defined by {\protect\cite{NumRec}}  : \\
$s_D = \left ( \frac{s_{\rm left}^2 + s_{\rm right}^2 }
{N_{\rm left} + N_{\rm right} -2} \right )^{1/2}
\left (\frac{1}{N_{\rm left}} + \frac{1}{N_{\rm right}} \right )^{1/2}$,
where $s_{\rm left}$ and $s_{\rm right}$ are the standard deviations
of the data to the left and to the right of the
pointer respectively, and  $N_{\rm left}$ and
$N_{\rm right}$ are the number of points to the left and to the right
of the pointer respectively.

\bibitem{siglevel}
The significance level ${\cal P}(\tau)$ of a possible cutting point
with $t_{\rm max}=\tau$ is defined as the probability of obtaining
the value $\tau$ or lower values within a random sequence:
${\cal P}(\tau)={\rm Prob}\left\{t_{\rm max}\le \tau \right\}$.
Note that this probability is not the same as the used for the standard
Student's test.
As we could not obtain ${\cal P}(\tau)$ in a closed analytical
form, we have developed an suitable approximation  by means of Monte
Carlo simulations.
${\cal P}(\tau) \approx
\left[ 1 - {\rm I}_{\frac{\nu}{\nu + \tau^2}} \left(\delta \nu,\delta \right )
\right]^\gamma $,
where $\gamma =4.19 ~ \ln N - 11.54$, $\delta = 0.40$,
$N$ is the size of the sequence or subsequence to be split, $\nu = N -2$,
the degrees of freedom, and $I_x(a,b)$ is the incomplete beta function
{\protect \cite{NumRec}}.



\bibitem{MIT} Data used in this study were provided, 
without cost, by PhysioNet (http://www.physionet.org/),
a public service of the Research Resource for Complex Physiologic
Signals, under a grant from the N.I.H./National
Center for Research Resources (P41 RR13622).  

\bibitem{night}
However, for the records during sleep, the distribution exhibits a crossover
at a characteristic segment length of 700 beats, which might be related to
the presence of sleep phases. This crossover indicates a smaller number
of segments with short length.

\bibitem{Peng93}
C.-K.~Peng {\sl et al.}, Chaos {\bf 5}, 82 (1995).


\bibitem{GenNoise}
H.A.~Makse {\sl et al.}, Phys. Rev. E {\bf 53}, 5445 (1996)


\bibitem{Ivanov96}
P.Ch.~Ivanov {\sl et al.}, Nature {\bf 383}, 323 (1996);
M.~Meyer {\sl et al.}, Integ. Physiol. and Behav. Sci. {\bf 33}, 344 (1998).



\end{references}
\end{document}